\documentclass[12pt]{article}
\begin{document}
\title{Why Do We See a Classical World?}
\author{Hartmann R\"omer \\Institute of Physics, University of Freiburg, Germany}
\date{}
\maketitle \pagenumbering{arabic}

\begin{abstract}
From a general abstract system theoretical perspective, a
quantum-like system description in the spirit of a generalized
Quantum Theory may appear to be simpler and more natural than a
classically inspired description. We investigate the reasons why
we nevertheless conceive ourselves embedded into a classically
structured world. Categorial, physical and pragmatic reasons are
proposed as explanations.
\end{abstract}
\section{Introduction}
The underlying world views of classical and quantum physics are
quite different. For contrasting purposes and neglecting
intermediate positions they might
be characterized as follows:\\
The world of classical physics is a realistic world of facts,
which exist independently of their observation and are registered
but not created by the act of measurement. \\
On the other hand, the world of Quantum Theory is a world of
potentialities, which, by the act of measurement,  are elevated to
a factual status as measurement results. As compared to classical
physics, the role of the observer is not only a receptive,
registering but an active and in part creative one. Indeed, the
violation of Bell's inequalities \cite{Bell2004} strongly suggests
an exclusion of local realism in the spirit of classical physics
and the Kochen-Specker theorem \cite{KS1967, Redhead} is an
obstruction for any realistic hidden variable theory with
non-contextual observables.
\\
In our everyday world we are used and inclined to consider the
classical world view as the view of common sense, whereas quantum
physics looks like a rather extravagant  view, admittedly imposed
by experimental facts but emerging only lately and being mainly
confined to the notoriously strange microphysical world.\\
In this note, we shall present evidence that quantum features of
the world are much more widespread and natural than suggested by
current common sense, in fact to such an extent that one may
wonder about the reasons for the strong favoring of the
classical view.\\
For what follows it is essential to realize that the world is not
directly given to us as such but only as and as far as it appears
to us on our inner screen. (Using a common philosophical term we
refer to this as to the \emph{phenomenal character of the world}.)
Probably, almost everybody will subscribe to this apparently
trivial statement, but, taken seriously, it leads to far reaching
consequences. The question is about the relationship between the
phenomenal and the "real" world. Naive realism asserts that the
world essentially appears to us as it really is. In the
terminology of Thomas Metzinger \cite{Metzinger2003} na\"\i{}ve
realism employs a \emph{transparent model}: We are modelling
creatures, creating representations of the outer world, of our
body and also higher order representations of our cognitive
system. A model is called \emph{opaque}, if it is recognizable as
a representation and \emph{transparent} (invisible), if its
representational character is not manifest and if, hence, the
representation is identified with the represented entity.\\
A reflection about the foundations of Quantum Theory and physics
in general must contain an investigation of the the prerequisites
given by the basics of the human mode of existence and
cognition, which are prior to any physical theory or act of measurement.\\
It is safe to say that the classical world view is closer to the
strong assumption of na\"\i{}ve realism than the quantum view,
which, attributing an active constitutive role to the observer, is
more aware of the phenomenal character of the world and, in a way,
more
cautious.\\
Caution and methodological prudence are no logically cogent reason
for a widespread "ontophobic" attitude of contemporary philosophy,
an abstention from any kind of ontological commitment in favor of
phenomenal, existential, language or discourse analytical
approaches. Later on we shall see that our cognitive system
strongly urges if not compels us to build at least tentative
ontological scenarios, for instance classically realistically
inspired ones as for some interpretations of
Bohmian mechanics \cite{BH}, or scenarios of quantum type.\\
Early on from the advent of quantum mechanics Niels Bohr was
convinced that the quantum theoretical figure of complementarity
was of universal significance far beyond the realm of physics.
Speculation along this line never ceased \cite{WR2000, GA2008}. In
particular Wolfgang Pauli pointed out the possibly universal
importance of quantum-like entanglement \cite{Pauli-Jung,
RecastingReality}. The idea of quantum reality gained unfortunate
popularity in esoteric circles but it was also followed in a
serious and formally well controlled way  \cite{Aertsetal}.
Indeed, a quantum analogue structure may be suspected to be
realized, whenever the order of
successive observations/measurements matters.\\
A world of strict quantum-like constitution would be a world of
potentialities. It would show a strongly phenomenal character,
because it would be an appearing world whenever a measurement
result becomes factual for an observer. From a less observer
centered point of view and using a philosophical term, such a
world might also be called a \emph{worlding} (German:
"weltend") world.\\
Assuming that the significance of quantum-like structural features
beyond the realm of physics in the narrow sense were a plain
direct effect of quantum physics would amount to an extreme
physical reductionism of very low plausibility. Rather one should
look for structural isomorphisms with quantum physics.  In
general, formal work on wider applicability of Quantum Theory
sought to employ the full quantum theoretical formalism to non
physical situations. An alternative is the isolation and
formalization of a conceptual core of Quantum Theory followed by
an investigation of the extended applicability of the resulting
generalized scheme. This has been undertaken under the name of
\emph{"Weak Quantum Theory"} or \emph{"Generalized Quantum
Theory"} \cite{ARW, AFR2006, FR2010}, which we are going the
describe in the next section.
\section{Generalized Quantum Theory}
Weak Quantum Theory \cite{ARW, AFR2006} arose from an axiomatic
formulation of physical Quantum Theory by leaving out all features
which seemed to be special for physical systems. The term "Weak
Quantum Theory" was chosen because the resulting system of axioms
is weaker than quantum physics. It is of course stronger in as
much as it has a wider range of applicability. In order to avoid
misunderstandings we now prefer the term "Generalized Quantum
Theory" (GQT). In order to make this presentation reasonably self
sustained we here repeat a short account of the vital structural
features of GQT to which we can refer in the sequel.
For recent developments and applications see \cite{FR2010}. \\
\vspace{0.2cm}
The following notions are taken over from quantum physics:\\
\emph{System}: A system is anything which can be (imagined to be)
isolated from the rest of the world and be subject to an
investigation. A system can be as general as an object or a school
of art together with all persons involved in production and
interpretation. Unlike the situation in, e.g., Classical mechanics
the identification of a system is not always a trivial procedure
but sometimes a creative act. In many cases it is possible to
define
\emph{subsystems} inside a system%
\\[0.2cm]
\emph{State}: A system must have the capacity to reside in
different states without losing its identity as a system. One may
differentiate between \emph{pure states}, which correspond to
maximal possible knowledge of the system and \emph{mixed states}
corresponding to
incomplete knowledge. \\[0.2cm]
\emph{Observable}: An observable corresponds to a feature of a
system, which can be investigated in a more or less meaningful
way. \emph{Global observables} pertain to the system as a whole,
\emph{local observables} pertain to subsystems. In the above
mentioned example systems, observables may correspond to esthetic
investigations for systems of (schools of) art.
\\[0.2cm]
\emph{Measurement}: Doing a measurement of an observable $A$ means
performing the investigation which belongs to the observable $A$
and arriving at a result $a$, which can claim factual validity.
What factual validity means depends on the system: Validity of a
measurement result for a system of physics, internal conviction
for self observation, consensus for groups of human beings. The
result of the measurement of $A$ will in general depend on the
state $z$ of the system before the measurement but will not be
completely determined by it. \vspace{0.4cm}

Moreover, to every observable $A$ we associate its
\emph{spectrum}, a set Spec\,$A$, which is just the set of all
possible measurement results of $A$.
Immediately after a measurement of an observable $A$ with result a in Spec\,$%
A$, the system will be in an \emph{eigenstate} $z_a$ of the
observable $A$ with \emph{eigenvalue} $a$. The eigenstate $z_a$ is
a state, for which an immediate repetition of the measurement of
the same observable $A$ will again yield the same result $a$ with
certainty, and after this repeated measurement the system will
still be in the same state $z_a$. This property, which is also
crucial in quantum physics justifies the terminology ``eigenstate
of an observable $A$'' for $z_a$ and ``eigenvalue'' for the result
$a$. We emphasize that this is an idealized description of a
measurement process abstracting from its detailed temporal structure.\\
Two observables $A$ and $B$ are called \emph{complementary}, if
the corresponding measurements are not interchangeable. This means
that the state of the system depends on the order in which the
measurement results, say $a$ and $b$, were obtained. If the last
measurement was a measurement of $A$, the system will end up in an
eigenstate $z_a$ of $A$, and if the last measurement was a
measurement of $B$, an eigenstate $z_b$ will result eventually.
For complementary observables $A$ and $B$ there will be at least
some eigenvalue, say $a$, of one of the observables for which no
common eigenstate $z_{ab}$ of both observables exists. This means
that it is not generally possible to ascribe sharp values to the
complementary observables $A$ and $B$, although both of them may
be equally important for the description of the system. This is
the essence of quantum theoretical complementarity which is well
defined also for GQT. \\
Non complementary observables, for which the order of measurement
does not matter, are called \emph{compatible}. After the
measurement of compatible observables $A$ and $B$ with results $a$
and $b$, the system will be in the same common eigenstate $z_{ab}$
of $A$ and $B$ irrespective of the order in which the measurements
were performed.\\
\emph{Entanglement} can also be defined in the framework of
Generalized Quantum Theory \cite{ARW, AFR2006, FR2010,
Roemer2011}. It may and will show up under the following
conditions:\\
\begin{enumerate}
\item Subsystems can be identified within the system such that
local observables pertaining to different subsystems are
compatible.

\item There is a global observable of the total system, which is
complementary to local observables of the subsystems.

\item The system is in an \emph{entangled state} for instance in
an eigenstate of the above mentioned global observable and not an
eigenstate of the local observables.
\end{enumerate}

Given these conditions, the measured values of the local
observables will be uncertain because of the complementarity of
the global and the local observables. However, so-called
\emph{entanglement correlations} will be observed between the
measured values of the local observables pertaining to different
subsystems. These correlations are non-local and instantaneous.\\
Comparing Generalized with physical Quantum Theory the following
vital differences are worth noticing: \vspace{0.4cm}
\begin{itemize}
\item In GQT there is no quantity like Planck's constant
controlling the degree of complementarity of observables. Thus,
contrary to physical Quantum Theory, where quantum effects are
essentially restricted to the microscopic regime, macroscopic
quantum-like effects in GQT are to be expected.

\item At least in its minimal version described here, GQT contains
no direct reference to time or dynamics.

\item In its minimal version, GQT does not ascribe quantified
probabilities to the outcomes of measurements of an observable $A$
in a given state $z$. Indeed, to give just one example, for
esthetic observables quantified probabilities seem to be
inappropriate from the outset. What rather remains are modal
logical qualifications like ``impossible'', ``possible'' and
``certain''. Related to the absence of quantified observables, the
set of states in GQT is in general not modelled by a linear
Hilbert space. Moreover, no addition of observables (operationally
difficult to access even in quantum physics) is defined in GQT.

\item Related to this, GQT in its minimal form provides no basis
for the derivation of inequalities of Bell's type for measurement
probabilities, which allow for the conclusion that the
indeterminacies of measurement values are of an intrinsic ontic
nature rather than an epistemic lack of knowledge.  In many (but
not all) applications of GQT indeterminacies may be epistemic and
due to incomplete knowledge of the full state or uncontrollable
perturbations by outside influences or by the process of
measurement. Notice that complementarity in the sense of GQT may
even occur in coarse grained classical dynamical systems \cite
{GA2006,GA2008}.
\end{itemize}
For some applications (see, e.g., \cite{AFR2004, ABFKR,
AFR2008,AR2012}, ) one may want to enrich the above described
minimal scheme of GQT by adding further structure, e.g., an
underlying Hilbert space
structure for the states. \vspace{0.4cm} \\
We should stress here that for very general systems like the above
mentioned schools of art, observables are not so directly given by
the system and read off from it like many mechanical observables.
On the contrary, as already suggested by the name of an
``observable'', the identification of an observable may be a
highly creative act of the observer, which will be essentially
determined by his horizon of questions and expectations. This
marks a decidedly epistemic trait of the notion of observables in
GQT even more so than in quantum physics. Moreover, the horizon of
the observer will change, not the least as a result of his
previous observations adding to the open and dynamical character
of the set of observables. What has just been said about
observables also applies to \emph{partitioning} a system into
subsystems. In fact, partitioning is achieved by means of
\emph{partition observables} whose different values differentiate
between the subsystems. In general, subsystems do not preexist in
a na\"\i{}ve way but are in a sense created in the constitutive
act of their identification.\\
Quantum-like phenomena like complementarity in the sense of GQT
may be expected whenever "measurement" operations change system
states and are not commutable. Such  situations should abound in
cognitive science and in everyday life. They apply in a
paradigmatic way to the human mind as seen from a first person
perspective, because the state of mind will invariably be changed
by the very act of its conscious realization. Human communities
provide another important field of possible applications of GQT.
Detailed empirical investigations of quantum features in
psychological systems have been performed for bistable perception
\cite{AFR2004, ABFKR, AFR2008}, decision processes, semantic
networks, learning and order effects in questionnaires
\cite{AR2012}. (See
\cite{QuantumConsciousness} for further information.)\\
From the general system theoretic point of view adopted in our
account of GQT and also from everyday experience, classical as
opposed to quantum-like systems should be a rather special and
rare case. They correspond to systems without complementarities:
All measurement operations commute without limitation and reveal
an underlying objective reality essentially untouched by the
measurements. This is a very strong assumption and a quantal world
view in the sense of GQT looks quite natural and suggested by
ontological parsimony. The natural and to some extent even a
priori character of quantum structure is clearly pointed out by M.
Bitbol. (See \cite{Bitbol} and references therein.) Asking for the
reasons why nevertheless a classical world view is widely favored
seems to be a legitimate
task.\\
\section{Fundamentals of the Mode of Human Existence}
Any reflection about the phenomenal character of the world
requires a detailed analysis of the mode of human existence as a
conscious being. This has been a main subject of philosophy since
the second half of the 19th century in particular of its
phenomenological line. Of course, in this study we can in no way
do justice to the vast body of work and thought done along this
line associated to prominent names like Franz Brentano, Edmund
Husserl, Martin Heidegger or Jean-Paul Sartre. For a deep and
comprehensive account see \cite{Prauss2006}. For our purposes, it
must suffice to point out a few constitutive characteristic basics
of human existence emerging from its
analysis:\\
\vspace{2mm} \centerline{ a) The figure of oppositeness \vspace{2mm}}\\
 Man always experiences his world as an observer, set
apart from and to some extent opposed to the object of his
attention. Ernst Tugendhat \cite{TugendhatDeutsch,
TugendhatSpanisch, TugendhatItalienisch} from the position of
analytic philosophy refers to this basic human existential as to
the \emph{"egocentricity"} of man as an "I-sayer". In quantum
physics the separation between observer and observed system is
known as the Heisenberg cut, which is movable but not removable.
In our more general framework we shall talk about the
\emph{epistemic cut}: Every cognition of a form accessible to us
is the cognition of someone about something. The location of the
epistemic cut may change depending on whether attention is
directed to an object outside or introspectively inside to the own
state of mind, but the epistemic cut never
disappears altogether.\vspace{4mm}\\
 \centerline{ b) Temporality \vspace{2mm}}\\
 Man's mode of existence is inescapably time-bounded.
The world appears to us not in the form of a simultaneous
panoramic picture but rather in the form of a movie: A narrow
window of a "now" is shifted over our reality giving a free direct
view only over an ever-changing small part of it. This internal
mental time is called by Mc Taggart \cite{McTaggart} an
\emph{A-Time}, which is characterized by the existence of a
privileged instance of a "now" and by its directedness towards a
future. In strong contrast to this, the outer time of physics is
what Mc Taggart calls a \emph{B-Time}, a scale time without a
privileged "now" and not necessarily directed. For the physical
origin of time directedness see \cite{Zeh}. More about the
difficult problem of the relationship between inner and outer time
in the framework of GQT may be found in \cite{Roemer2012} and
\cite{Roemer2004}. On an increasingly fundamental level of
physics, proceeding from Newtonian Mechanics to Special and
General Relativity Theory, physical B-Time shows a tendency to
become more and more similar to space and eventually to fade away
as a fundamental notion if quantum effects of space-time and very
strong gravitational fields are considered. (See \cite{Roemer2012}
and references therein.) However, internal A-Time persists and
leaves deep traces in thermodynamics via the close relationship
between the thermodynamic time arrow \cite{Zeh} and the so-called
psychological time arrow and, as we shall see in a moment also in
Quantum Theory. The two basic
existentials to be mentioned next are closely related to temporality.\\
\vspace{2mm} \centerline{ c) Facticity \vspace{2mm}}  We conceive
ourselves as living in a world of facts. The feeling of certainty
of a visual perception and the immediate presence in introspection
all carry an inexorable imprint of facticity. The "now" is located
in the heart of both temporality
and facticity. Facts underly Boolean Logic.\\
\vspace{2mm} \centerline{ d) Causality and freedom \vspace{2mm} }\\
Causality and freedom of action are both offshoots of
the same common root of a developed temporality unfolded into
past, presence and future. Rather than being in an exclusively
contradictory relationship they rely on each other, because
freedom is only possible if actions have foreseeable consequences
and causality can only be seen if there is freedom in the choice
of causes and initial conditions.\\
\vspace{2mm} \centerline{ e) Agentivity \vspace{2mm}}\\
 In our existence we experience ourself as agents, who
actively steer the focus of their attention and their bodily
motions. Planning, worrying and procuring are our future directed
activities and attitudes. In this context it is also worth remembering
that "factum" literally means "made".\\
\vspace{2mm} \centerline{ f) Emotionality \vspace{2mm}}\\
 This study is centered around the cognitional
activity of man. Nevertheless, it should be kept in mind that
emotions color all our perceptions and cognitions. We are
continuously assessing and judging. Emotions guide our will and
intentions, are constitutive for our personality and lie at the
basis of our creativity.\\

We already saw that (Generalized) Quantum Theory, more so than
classical theory, takes into account the phenomenal character of
our world. So, we should ask ourselves, whether the basic
categorical existentials enumerated above are reflected in the
structure of GQT. This,
indeed, turns out to be true to a large extent:\\
a) The structures of \emph{oppositeness and epistemic cut} are
deeply rooted in the distinction between system and observer as
well as in the central role attributed to measurement. Observables
neither exclusively pertain to the observer nor to the observed
system but
could be said to be located astride of the epistemic cut.\\
b) \emph{Temporality} leaves a subtle trace in the vital
importance of the (temporal) order of measurements. If observables
$A$ and $B$ can be composed, their composition $AB$ means
$A$\emph{after} $B$. In addition, the facticity of measurement
results, mentioned under point c), enters via the "now" of human
A-Time.  \\
c) \emph{Facticity} is strongly present in the factual validity of
measurement results. In a quantum picture of the world, a quantum
state before measurement describes a world of potentialities or,
more precisely, of timelessly extended simultaneity rather than
factual localization in a "now". From this point of view, every
completed measurement corresponds to an inroad of a classical
world into a quantum world. \\
d) and e) become apparent in GQT in the planning and execution of
experiments, and in the choice of observables to be measured. They
may
also be formalized in dynamical equations of motion. \\
f) Beyond its general great importance, \emph{emotionality} does
not play any special role in GQT, which is essentially a theory of
cognition. Moreover, and for good reasons, science strives for
emotional neutrality. However, systems of GQT may possess
emotional observables concerning e.g. mood, contention,
pleasantness, esthetic or moral value. Such variables pertain to
the cognitive, assessing component of emotions, which after all is almost never missing.\\
The above-mentioned categorical existentials are to some extent
suggestive of a classical world view.  \emph{Evolutionary
epistemology} \cite{Popper, Vollmer} asserts that our cognitional
system, which is based on these existentials adaptively arose by
Darwinian evolution: mutation and selection. Comparison with other
forms of life and with older pre-lingual stages of man shows
beyond any doubt that an evolution indeed occurred. It is also
clear that our cognitional system should  not jeopardize our
chances of survival. On the other hand, one should not overlook
some problematic features of evolutionary epistemology, at least
in its most popular interpretation:
\begin{itemize}
\item The environment, to which adaptation of the cognitional
system has to proceed is normally conceived as being of classical
type, often even identified with a classical physical system.
Quantum notions are usually not assumed to be relevant. This
classical environment is normally considered to be rigid and not
subject to evolution, at least as long as cultural evolution does
not become topical. Evolution time is identified with a directed
physical time of B-type in the sense of Mc Taggart
\cite{McTaggart}. In addition, evolutionary epistemology often
relies on a strong classical background materialism and
reductionism. This implies the danger of a gross underestimation
of the phenomenal character of our world. The world view of
classical physics arises from a particular modelization of the
world. As already mentioned in the Introduction, this is not
completely illegitimate as a tentative ontological scenario.
However, in a na\"\i{}ve realistic world view this model has
become completely transparent and a certain degree of opaqueness
seems to be desirable. \item Even if we take the correctness of
the central hypotheses of evolutionary epistemology for granted,
the survival success of the evolved cognitive system in no way
guarantees the ontological validity of the emerging culture
dependent world view, let alone of reductive classical
materialism. On the contrary, there are many examples, in
particular in cultural history demonstrating that the
evolutionally more viable view is not necessary the more correct
one.
\end{itemize}
\section{Excursus: Language}
Language is an inseparable part of our human psychic endowment.
So, we should not be surprised to find the basic existentials of
the previous section in human language. We shall demonstrate this
for 1) Facticity, 2) Temporality and 3) Agentivity: \\
1) Facticity\\
Facticity is reflected in what is called the \emph{propositional
character of language} \cite{TugendhatDeutsch, TugendhatSpanisch,
TugendhatItalienisch}: A normal uttering in human language is
either a clause of statement or question. The former directly
claims facticity, and the latter asks about facticity. The only
exceptions are exclamations and imperative sentences. Both are
archaic and syntactically isolated. Imperatives are typically the
most simple forms of the verb.\\
2) Temporality\\
Temporality is met in human languages in various forms
\begin{itemize}
\item It is manifest in the threefold temporal sequentiality of
language in sounds, words and sentences. \item Reference to time
is expressed in the verb in many ways. \emph{Tenses} express
temporal location with respect to the speaker (ex: "He wrote") and
sometimes also with respect to the reported action (ex: "He had
written"). \emph{Modes of action} are related to the lexical
meaning of a verb and describe the temporal form of the action
(durative, ingressive, iterative, punctual,...) and
\emph{aspects}, which are of key importance e.g. in Slavic
languages, are forms of the verb allowing to express whether the
speaker wants to report on the action as ongoing or as a completed
entity \cite{Aitzetmueller}. (English ex: "He was writing a
letter" vs "He wrote a letter")
\end{itemize}
3) Agentivity\\
The default attitude whether a speaker understands himself
primarily as (a) an acting or as (b) an experiencing being differs
between various languages. It has several linguistic reflexes
which show a tendency to be correlated:
\begin{itemize}
\item Most European  languages  favor attitude (a). For these
languages the main distinction is between tenses, which is
morphologically most clearly expressed is the distinction between
past and non-past (present/future), because it coincides with the
distinction between "non influenciable" and "influenciable". For
attitude (b) the main distinction tends to be between future and
non future (presence/past), which corresponds to the distinction
between invisible and visible. Eskimo languages are an example for
this state of affairs \cite{Holst}. \item European peoples
normally conceive the future as approaching us from the front and
receding to the past which lies behind us. This is in line with an
active attitude (a), which considers the future as something to be
faced and influenced. The converse view, in accordance with
attitude (b), for which the invisible future approaches from the
back side and turns into the visible presence and past in front of
us has been observed in Babylonian \cite{ZeitBabylon} and Aymara
\cite{ZeitAymara}. For instance, in Babylonian future literally
means "lying in the back" and past "lying in front". Aymara
speakers point backwards when referring to the future. \item The
difference between the active attitude (a) and the receptive
attitude (b) may also be mirrored in a preference for a
\emph{accusativic} and \emph{ergativic} \cite{Dixon, Holst}
sentence structure. Let us briefly explain this: Intransitive
verbs (ex: "to sit") have only one participant, the \emph{subject}
(S) (ex: "Peter (S) is sitting"). The subject  normally stands in
the most simple unmarked case, the \emph{nominative}. Transitive
verbs (ex: "to hit") have (at least) two participants, the
\emph{actor} (A) and the \emph{experiencer} (E) (ex: "Peter (A)
hits the ball (E)"). Almost all European languages except Basque
employ an accusativic sentence structure for transitive verbs: The
actor (A) of a transitive verb stands in the nominative case just
like the subject (S) of the intransitive verb, whereas the
experiencer (E) stands in a different case, the \emph{accusative}.
(In English, where nominative and accusative are morphologically
differentiated only for pronouns, both (S) and (A) stand before
the verb and (E) behind the verb.) This parallel treatment of (S)
and (A) signals an active attitude placing the actor in a
privileged primary position. \\Basque and many languages outside
Europe (Caucasian languages, Eskimo languages, Maya languages,
Australian aboriginal languages, Chukotian languages,...) choose a
different sentence construction for transitive verbs: The
syntactic position of (E) runs in parallel with (S), whereas  (A)
stands in a different case called \emph{ergative}. Here, the
pivotal position is occupied by the receptive experiencer (E). The
ergativic sentence construction is somewhat similar to the passive
construction in European languages (ex: "The ball (E) is hit by
Peter (A)"). However, the European passive only arises by an
additional transformation of an active sentence and the accusative
construction is the default. In ergative languages the ergative
structure is the default. (Indeed, many ergative languages have an
"antipassive" transformation yielding an analogue of the normal
sentence construction of accusative languages.) Let us finally
mention that many languages (e.g. Georgian and Sumerian) have what
is called an \emph{split ergative} structure: Depending on the
tense of the transitive verb an accusative or ergative
construction is applied. Not surprisingly, the ergative sentence
structure is favored in the past tense, because an action in the
past cannot be really performed but only reported or imagined.
\end{itemize}
\section{Why Classical?}
We have argued that in many respects a quantum-like world view
seems to be more natural and ontologically parsimonious. Moreover,
our introspective world as well as much of our outside world, at
least on closer inspection, makes a quantum-like impression. In
what follows, we shall give (A) categorial, (B) physical and (C)
pragmatic reasons for our strong inclination to conceive ourselves
as living in a classical world. None of them is completely cogent.
After all, by a special intellectual effort, man has proved to be
capable to device a quantum-like world view and even to get
acquainted to it to some extent. But taking all these reasons
together, our predilection for a classical world view becomes
almost irresistible, at least for everyday life. \vspace{3mm}\\
A) We already mentioned that the basic categorical existentials of
section 3  rather suggest a classical world view. This in
particular applies to the existencial of facticity. Our world, as
we experience it, is inescapably fact like, which is also
reflected in the propositional character of our language. From our
very nature we have a deeply rooted tendency to be na\"\i{}ve
realists unhesitatingly taking the representations on our internal
screen as the real world. Metzinger \cite{Metzinger2003} asserts
that transparent models are evolutionally favored. In fact, in
view of an approaching predator it would be a waste of time and
energy for life saving reaction to realize the representational
character of its appearance on our inner stage. On a higher level,
we are naturally inclined to ontologize what on closer scrutiny
could only be granted a phenomenal status. This predilection for
ontological scenarios is an inseparable part of our mental
endowment and of our culture. We already pointed out that an
ontophobic ascetism may be barren. Ontologization is invaluable
for understanding and orientation in our world, as long as some
degree of fluidity is preserved, which sometimes allows us to look
behind the screen and to correct inappropriate one-sidedness,
petrifaction and sclerotization. In particular, this kind of
fluidity allows for the setup of both quantum theoretically and
classically inspired world scenarios.
\vspace{3mm} \\
 B) The macroscopic validity of Classical Mechanics is
often invoked as the reason for the classical appearance of the
world.  In the macroscopic regime, to which Classical mechanics
applies, quantum uncertainties are normally invisible because of
the smallness of Planck's constant $\hbar$. Moreover, by the
\textit{quantum Zeno effect} \cite{MS1977}, repeated measuring and
monitoring of the system will prevent an uncontrolled growth  of
uncertainties. From a fundamental point of view, the macroscopic
classical limit of Quantum Theory and the measurement process as
an interaction with a macroscopic measurement device are not
completely understood in quantum physical terms, at least not for
individual systems, rather than ensembles. Decoherence theory
\cite{GJKKSZ} goes an important step in this direction. It
explains how normal unitary time evolution of pure states of
macroscopic systems coupled to an environment leads to states
which, by local measurements on the system, are indistinguishable
from mixed states. The decoherence time needed to reach such
states quickly decreases with the sizes of the system and the
environment and is typically very small. What is not described by
decoherence theory is the collapse of the wave function, the
transition from potentiality to measured facticity, which does not
correspond to a unitary evolution in time. Indeed, for individual
physical systems instead of ensembles there is so far no
description of the collapse in terms of normal unmodified Quantum
Theory. This may be interpreted as a hint that measurement is not
exclusively to be understood as a physical process but as an act
of cognition, which is, of course, accompanied by a physical
process on a physical substrate but not to be identified with this
physical process. In fact, no clear physical criterium seems to be
in sight qualifying a physical process as a measurement process or
as an act of cognition. This remark about a possible non physical
but cognitive nature of the measurement process applies to GQT
even more than to quantum physics. A physical analysis of a
measurement process is important even if it does not capture all
its cognitional aspects. The situation presents itself as follows:
The requirement of the possibility of cognition is of course
logically prior to any kind of physics and physical measurement.
The result of an investigation of the physical process
accompanying an act of cognition and measurement must be
consistent with this possibility. The Quantum Theory of the
measurement process meets this requirement very well. The
measurement process is described by a quantum theoretical system
containing the measured system $S$, the measuring device $M$ and
possibly some environment $E$. An entangled state evolves by
unitary time evolution, which, by reduction to the measuring
device $M$ yields a mixed state of $M$ reproducing exactly the
probabilities of measurement results for $S$ predicted by Quantum
Theory for a state $\rho $ of $S$ before measurement. The same
probabilities are also obtained  by applying the measurement of
$S$ in  the state $\rho '$ of S arising by decoherence theory
after reducing an evolved entangled state $\rho ''$ of $S+M(+E)$
to $S$.
\vspace{3mm}\\
Given the macroscopic validity of Classical mechanics, we should
not forget that Classical Mechanics only describes a narrow and
highly idealized sector of the world in which we find ourselves
living. As already mentioned, other important parts of it,
including our inner and social world, are rather quantum-like
constituted in the sense of GQT. So, the hint to Classical
Mechanics does not really answer our original question but rather
rephrases it in the form: Why do we attribute so much importance
to Classical Mechanics in the formation of our world model? A
categorial reason for this inclination has already been given
under (A). There are other logically not completely unrelated
reasons for favoring a classical world view: \\
 C1) Man in his temporal mode of existence has good
reasons to keep to the more stable and reliable features of his
physical and social environment. In the material world, a bow
spanned and pointed the same way must produce the same shot and a
leap done with the same force must carry over the same distance.
The necessary stability of a human society is based on a common
stock of accepted facts and values and a collection of compatible
observables and of histories whose consistency \cite{Griffiths} is
generally acknowledged. A cultural habitat of (floating) islands
of stability is woven as a result of continuous collective work.
(This comparison comes from a visit of the Uru-Chipaya tribe, who
really lives on floating islands on lake Titicaca built from reed
and continuously enlarged and repaired also by incorporation of
waste.) The subtle and impressive building of classical natural
science is a monumental example of probably the largest consistent
structure of our time. Historiography and belief systems build
other islands. Hans Primas \cite{Primas2007} talks about
\emph{partially Boolean Systems}. Our cultural activity tries to
extend them as much as possible. Consistency between different
islands and sometimes even inside the islands cannot always be
achieved, if complementarity is really a general constitutive
feature of the world. For the sake of cohesion of society it is
natural not to stress but rather to suppress such inconsistencies
and anomalies. All this leads to
the stabilization of a world view of predominantly classical type.\vspace{3mm}\\
C2) All kind of information is factual, even information about
Quantum Theory. In our life we are swamped with (hard) facts,
which peremptorily call for attention, respect and action. The
inevitability of death is a particularly grim example of impending
factuality. The possibility to store and accumulate facts as
documents further adds to their overwhelming dominance.\vspace{3mm}\\
C3) In a world of surprises and unpredictability man tends to
explain uncertainties by lack of knowledge or understanding. This
suggests a classical background model of the world, which is
difficult empirically to tell apart from a quantum-like model. The
key paradigm of unpredictability is the autonomous behavior of
personal beings. Quite naturally in earlier stages of mankind
animistic world models prevailed and soothing and reconciliating
strategies were largely employed to influence potentially
dangerous or helpful personal instances. Even for the rather
quantum-like internal world intuitions and dreams were widely
interpreted as messages from outside intelligences. The
development proceeded in the direction of successively
substituting personal agents by "natural" ones, which promised a
higher degree of control and understanding. A culmination of this
development is marked by the success of deterministic Classical
Mechanics together with a program of replacing all spiritual
aspects of the world by physical reductionism. In addition,
classical logic seemed to imply a classical world view. (In fact,
also Quantum Theory can be formulated with classical logic.)\vspace{3mm}\\
Finally, we should mention that also in GQT a quantum Zeno effect
\cite{MS1977, AFR2004, ABFKR, AFR2008} strengthens the facticity
of measurement results, which can be stabilized and held fixed  by
continuous observation and sufficiently frequent repetition of a
measurement.
\section{Concluding Remarks}
Although many factors, including our categorical framework, urge
us to adopt a classical world view, this tendency is not an
inescapable fate. Man at least has the capability to reflect on
his
categorial endowment, to question it  and to try a  glimpse behind this curtain.\\
We already mentioned several times that large parts of our world
are organized in a quantum-like way, even if a classical
background model prevents us from acknowledging this explicitly
and suggests alternative terminologies and explanations. The human
mind and its products, the  internal and social world of human
beings are quantum reservations. \\ The simultaneous presence of
alternatives in a quantum state has an enormous creative
potential, which may very well be active in such highly creative
processes like formation of concepts, identification of systems,
detection of observables and also in social empathy and cultural
activities like poesy and fine arts. The notion of \emph{implicate
order} developed by D. Bohm and B. Hiley \cite{Bohm1980, BH} is
closely related to this creative potential. It would be surprising
if evolution had not made use of it, and the work on the
development of a quantum computer is an
endeavor to exploit it even technically.\\
 Moreover, the Quantum Theory of measurement teaches us that
measurement/cognition are realized by means of quantum
entanglement correlations. \\
There is another reason that the limitations imposed on us by the
framework of our categorical existentials are not unsurmountable:
Mankind is continuously striving to transcend its own categorical
framework. In fact, the very term of "existence" literally means
"stepping out". This tendency is already prepared in the phylogeny
of man and repeated in its ontogeny. The temporality of simple
animals strictly confines them to a narrow "now". The unfolding of
temporalty into present, past and future is an act of
emancipation. The possibility to re-present other instances of
time enormously widens the temporal screen. Planning, worrying and
freedom of action now become possible. \\Language enables symbolic
representations and an emancipation from blunt facts in a mode of
contrafactuality, in which the space of possibilities can be
freely explored. Under this perspective, the emergence of Quantum
Theory may be interpreted as a late highlight
in this emancipatory process. \\
Man also rebels against the limitation imposed by oppositeness and
the epistemic cut trying to see himself integrated and secured in
an all-comprising world. Seeking mystic unity
\cite{TugendhatDeutsch, TugendhatSpanisch, TugendhatItalienisch}
or strict mechanistic reductionism can be seen to stand for two
opposite extremal attempts to overcome the structure of an
individuum confronted to its world. Both of them tend to neglect
the phenomenal character of the world, which is taken into account
in a balanced and subtle way by Quantum Theory.


\end{document}